\documentclass[prl,twocolumn,twoside,showpacs,byrevtex,superscriptaddress]{revtex4}

\usepackage{currfile}

\usepackage{graphicx,epsfig,verbatim,enumerate}
\usepackage{amssymb,amsmath}
\usepackage{ifthen}
\newboolean{twocolswitch}

\newcommand{\sindex}[1]{}
\newcommand{\nindex}[1]{}

\newcommand{\www}[1]{\url{#1}}

\usepackage{lettrine}

\usepackage{color}

\setboolean{twocolswitch}{true}

\begin{document}

\title{
  
Vaporous Marketing: Uncovering Pervasive Electronic Cigarette Advertisements on Twitter

}

\author{
\firstname{Eric M.}
\surname{Clark}
}

\email{eclark@uvm.edu}

\affiliation{University of Vermont}
\affiliation{Department of Mathematics \&  Statistics}
\affiliation{Vermont Complex Systems Center}
\affiliation{Vermont Advanced Computing Core}
\affiliation{Computational Story Lab}
\affiliation{Department of Surgery}

\author{
\firstname{Chris A.}
\surname{Jones}
}
\affiliation{University of Vermont}l
\affiliation{Department of Surgery}
\affiliation{Global Health Economics Unit of the Vermont Center for Clinical and Translational Science}
\affiliation{Vermont Center for Behavior and Health}

\author{
\firstname{Jake Ryland}
\surname{Williams}
}
\affiliation{University of Vermont}
\affiliation{Department of Mathematics \&  Statistics}
\affiliation{Vermont Complex Systems Center}
\affiliation{Vermont Advanced Computing Core}
\affiliation{Computational Story Lab}

\author{
\firstname{Allison N.}
\surname{Kurti}
}
\affiliation{University of Vermont}
\affiliation{Vermont Center for Behavior and Health}

\author{
\firstname{Michell Craig}
\surname{Nortotsky}
}
\affiliation{University of Vermont}
\affiliation{Department of Surgery}

\author{
\firstname{Christopher M.}
\surname{Danforth}
}
\affiliation{University of Vermont}
\affiliation{Department of Mathematics \&  Statistics}
\affiliation{Vermont Complex Systems Center}
\affiliation{Vermont Advanced Computing Core}
\affiliation{Computational Story Lab}

\author{
\firstname{Peter Sheridan}
\surname{Dodds}
}
\affiliation{University of Vermont}
\affiliation{Department of Mathematics \&  Statistics}
\affiliation{Vermont Complex Systems Center}
\affiliation{Vermont Advanced Computing Core}
\affiliation{Computational Story Lab}

\date{\today}

\begin{abstract}
  
\hspace{-4mm}\textbf{Background}  \\
Twitter has become the``wild-west" of marketing and promotional strategies for advertisement agencies.  Electronic cigarettes have been heavily marketed across Twitter feeds, offering discounts, ``kid-friendly" flavors,  algorithmically generated false testimonials,  and free samples.  \\

\hspace{-5mm} \textbf{Methods} \\
All electronic cigarette keyword related tweets from a  10\% sample of Twitter spanning January 2012 through December 2014  (approximately 850,000 total tweets) were identified and categorized as Automated or Organic by combining a keyword classification and a  machine trained Human Detection algorithm.   A sentiment analysis using Hedonometrics was performed on Organic tweets to quantify the change in consumer sentiments over time.  Commercialized tweets were topically categorized with key phrasal pattern matching. \\

\hspace{-5mm} \textbf{Results} \\
The overwhelming majority (80\%) of tweets were classified as automated or promotional in nature.  The majority of these tweets were coded as commercialized (83.65\% in 2013),  up to 33\% of which offered discounts or free samples and appeared on over a billion twitter feeds as impressions.
The positivity of Organic (human) classified tweets has decreased over time  (5.84 in 2013 to 5.77 in 2014)  due to a relative increase in the negative words `ban', `tobacco', `doesn't', `drug',  `against', `poison', `tax' and a relative decrease in the positive words like `haha', `good', `cool'.  Automated tweets are more positive than organic (6.17 versus 5.84) due to a relative increase in the marketing words like `best', `win', `buy', `sale', `health', `discount'  and a relative decrease in negative words like `bad', `hate', `stupid', `don't'.\\  

\hspace{-5mm} \textbf{Conclusions}\\
 Due to the youth presence on Twitter and the clinical uncertainty of the long term health complications of electronic cigarette consumption, the protection of public health warrants scrutiny and potential regulation of social media marketing. \\
  
\end{abstract}

\maketitle

\section*{Introduction}
Electronic Nicotine Delivery Systems, or e-cigs, have become a popular alternative to traditional tobacco products. The vaporization technology present in e-cigarettes allows consumers to simulate tobacco smoking without igniting the carcinogens found in tobacco \cite{cobb2010novel}.  Survey methods have revealed widespread awareness of e-cigarette products \cite{zhu2013use,pearson2012cigarette}.   The health risks \cite{vansickel2010clinical,goniewicz2014levels,callahan2014electronic,kosmider2014carbonyl}, marketing regulations  \cite{trtchounian2011electronic}, and the potential of these devices as a form of nicotine replacement therapy \cite{kandra2014physicians,grana2014cigarettes,eissenberg2010electronic} are hotly debated  politically \cite{EuopeanVoiceBattleEUban} and investigated  clinically \cite{palazzolo2013electronic,avdalovic2012electronic}.
The CDC reports that more people in the US are addicted to nicotine than any other drug and that nicotine may be as addictive as heroin, cocaine, and alcohol \cite{centers2014health,nicoteneaddictive,policytobacco,us2010tobacco}.  Nicotine addiction is extremely difficult to quit, often requiring more than one attempt \cite{us2010tobacco,us2000reducing}, however nearly 70\% of smokers in the US want to quit \cite{centers2011quitting}.    Data mining can provide valuable insight into marketing strategies, varieties of e-cigarette brands, and their use by consumers \cite{kim2015using,yip2013mining,grana2014smoking,zhu2014four,kim2015using,aphinyanaphongs2016text}.
 
 \par

 Twitter, a mainstream social media outlet comprising over 230 million active accounts, provides a means to survey the popularity and sentiment of consumer opinions regarding e-cigarettes over time.  Individuals post tweets which are short text based messages restricted to 140 characters.   Using data mining techniques,  roughly $850,000$ tweets containing mentions of e-cigarettes were collected from a 10\% sample of Twitter's garden hose feed spanning from January 2012 though December 2014.   This analysis extends a preliminary study  \cite{huang2014cross}  which analyzed all e-cigarette related tweets spanning May through June 2012.
 
    As Twitter has become a mainstream social media outlet, it has become increasingly enticing for third parties to gamify the system by creating self-tweeting automated software to send messages  to organic (human) accounts as a means for personal gain and for influence manipulation \cite{EvilDataSci}. We recently introduced a classification algorithm that is based upon three linguistic attributes of an individual's tweets \cite{clark2015sifting}.  The algorithm analyzes the average hyperlink (URL) count per tweet,  the average pairwise dissimilarity between an individual's tweets, and the unique word introduction decay rate of an individual's tweets.
 
All tweets mentioning e-cigarettes were categorized using a two-tier classification process.   Tweets containing an abundance of marketing slang (`free trial', `starter kit', `coupon') are immediately categorized as automated.   All of the tweets from individuals that have mentioned an e-cigarette keyword are collected in order to classify the remaining tweets per individual as either organic or automated.   The machine learning classifier was trained on the natural linguistic cues from  human accounts to identify promotional and SPAM entities by exclusion.   

   The manipulative effects, agendas, and ecosystem of generalized social media marketing campaigns have been identified and extensively studied \cite{lee2013crowdturfers, ranganathunderstanding, wang2012serf}.  Other work, \cite{chu2012detecting}, has distinguished between purely automated accounts, or ``robots", and human assisted automated accounts referred to as ``cyborgs".  On Twitter, these campaigns have also been characterized using  Markov Random Fields to classify  accounts as either promotional or organic \cite{li2014detecting}.    This study was able to achieve very high classification accuracy, but was working under a much shorter time frame (1 month) and was trained on all relevant tweets authored within this time window.  Our study compiled a 10\% sample of tweets over a three-year period, so we relied on a classifier that was trained on smaller samples of tweets per individual.  

 The emotionally charged words that contribute to the positivity of various subsets of tweets from each category were quantitatively measured using hedonometrics  \cite{dodds2011temporal,dodds2014human}.  Outliers in both the positivity and frequency time-series distributions correspond to political debates regarding the regulation of e-cigarettes.  Recent studies\cite{dutra2014electronic,cho2011electronic,pepper2013adolescent,goniewicz2012electronic,wills2015risk}  report an alarmingly rapid increase in the youth awareness and consumption of  electronic cigarettes; a Michigan study found that the use of e-cigarettes surpass tobacco cigarettes among teens \cite{johnston2014monitoring}.  The CDC reports that ``the number of never-smoking youth increased three-fold from approximately 79,000 in 2011 to 263,000 in 2013" \cite{bunnell2014intentions}.  During this time-period there has also been a substantial (256\%) increase in youth exposure to electronic cigarette television marketing campaigns \cite{duke2014exposure}.  Due to the high youth presence on Twitter \cite{brenner201372} as well as the clinical uncertainty regarding the risks associated with e-cigarettes, understanding the effect of promotionally marketing vaporization products across social media should be immediately relevant to public health and policy makers.

\section*{Materials and Methods}
\subsection*{Data Collection}

    An exhaustive search from the 10\% ``garden hose" random sample of Twitter spanning  2012 through 2014  yielded approximately 850,000 tweets mentioning a keyword related to electronic cigarettes including: e(-)cig, e(-)cigarette, electronic cigarette, etc.  All tweets were tokenized by removing punctuation and performing a case insensitive pattern match for keywords.  Using time zone meta-data the tweets were converted into their local post time, in order for a more accurate ordinal sentiment analysis.  The language, reported by Twitter, and user features were also collected and analyzed.
    
\subsection*{Automation Classification}
  As reported in \cite{huang2014cross} there is a high prevalence of automation among e-cigarette related tweets.  Many of these messages were promotional in nature, offering discounted or free samples or advertising specific electronic cigarette paraphernalia (see Table 3).  A human detection algorithm defined and tested  in \cite{clark2015sifting} was implemented to classify  accounts as either automated or organic (human in nature).  All tweets from each individual appearing in our dataset were collected for the classifier.   For each individual, the average URL count, average tweet dissimilarity, and word introduction decay rate were calculated for the individuals with at least 25 sampled tweets. 
  
    \par
      The majority (94\%) of commercial e-cigarette tweets collected by \cite{huang2014cross} contain a hyperlink (URL).  The average URL count per tweet has been demonstrated to be a strong feature for detecting robotic accounts \cite{HumanBotCyborgTwitter,SocialHoneyPotProtecting,SocialHoneypotsMachineLearning}.  Many algorithmically generated tweets contain similar structures with minor character replacements and long chains of common substrings, as opposed to Organic content.  The Pairwise Tweet Dissimilarity of tweets $t_i,t_j$ from a particular individual was estimated by subtracting the length (number of characters) of the longest common subsequence, $|LCS(t_i,t_j)|$ from the length of both tweets, $|t_i|+|t_j|$ and normalizing by the total length of both tweets:
          $$ \text{D}(t_i,t_j) = \dfrac{ |t_i| + |t_j| - 2\cdot|LCS(t_i,t_j)|}{ |t_i|+|t_j|}. $$  
          
          \par 
          
             For example, given the two tweets: \\ $(t_1, t_2)$ = (I love tweeting, I love spamming).  Then $|t_1|$ = 16, $|t_2|$ = 15, $LCS(t_1,t_2) = |\text{I love }| = 7$ (including whitespace) and we calculate the pairwise tweet dissimilarity as: 
   $$D(t_1,t_2) = \dfrac{16 + 15 - 2\cdot 7 }{16+15} = \dfrac{17}{31}.   $$
   
   \par
          
          The average tweet dissimilarity of the individual was then estimated by finding the arithmetic mean of each individual's calculated pairwise tweet dissimilarity.  Since automated and promotional accounts have a structured and limited vocabulary,  the unique word introduction decay rate introduced in \cite{TxtMixing} serves as another useful attribute to detect automated accounts.  Using these attributes, the calibrated human detection algorithm, tested in \cite{clark2015sifting},  detected over 90\% of automated accounts from a mixed 1000 user sample with less than a 5\% false positive rate. 
  
   The Human Detection Algorithm was calibrated for a range of tweet sample sizes from hand classified Organic accounts.   Ordinal samples of collected  tweets from each account were binned into partitions of 25 ranging from 25 to a maximum of 500 tweets.  Table 1 below lists the number of automated and organic classified accounts per year.   Individuals with less than 25 sampled tweets were not classified with the detection algorithm.

  \begin{table}[Ht]\caption{Human Detection Twitter Account Classification}
\begin{tabular}{|cccc|}
\cline{1-4}
 Year & Automated & Organic &  Unclassified*\\
 \hline
  2012 & 12,715 & 12,052 & 19,512 \\
  2013 & 64,874  & 59,376 & 120,142 \\
  2014 & 54,033 & 63,289 &  48,528 \\
\hline
\end{tabular}
\\

*account had less than 25 sampled tweets
\label{HDuserClass}
\end{table}

   To benchmark the accuracy of the detection algorithm on this sample of tweets, a random sample of 500 accounts algorithmically classified as automatons and 500 classified as Organic were hand classified.  In Figure 1 below, features of each of these 1000 sampled individuals are plotted in three dimensions.  Organic features (green) are densely distributed, while the automated features (red points) are more dispersed.  The black lines illustrates the organic feature cutoff for the classifier; individuals with features falling outside of the box are classified as automatons.  On this sampled set of accounts, the classification algorithm exhibited a 94.6\% True Positive rate with a 12.9\% False Positive Rate.
        
    \begin{figure}[Ht]
      \includegraphics[width=.5\textwidth]{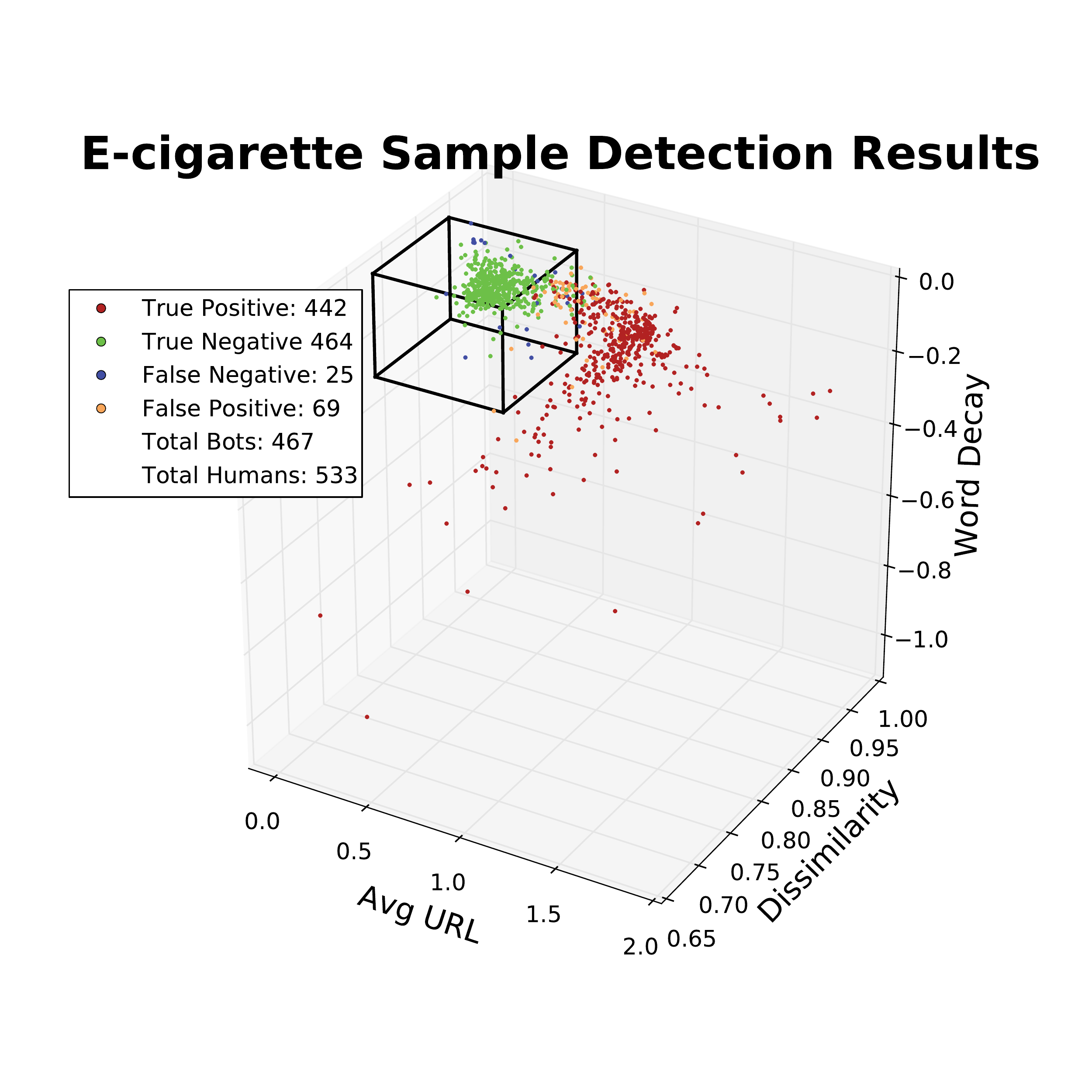}                        
 \caption{  Tweets from a random sample of 500 organic classified and 500 automated classified accounts were hand coded to gauge the accuracy of the detection algorithm.  The feature set of each sampled individual is plotted in three dimensions.  The traced box indicate the organic feature cutoff.  True Positives (red) are correctly identified automatons, True Negatives (green) are correctly identified Humans,  False Negatives (blue) are automatons classified as humans and False Positives (orange) are humans classified as automatons. }
\label{fig:RobotCloud}
 \end{figure}

 \subsection*{ Categorization by Topics}
   
     Tweets with at least 3 advertising jargon references (e.g. coupon, starter kit, free trial)  were immediately classified as automated.  All posts from users with at least 10 marketing classified tweets were also flagged as automated.  As noted in \cite{huang2014cross}, some Organic users could retweet promotional content for rewards (e.g. winning free samples or discounts).  All of these tweets were still classified as automated, but the user was not flagged as such.  The remaining tweets were classified as either automated or organic by the human detection algorithm.  Posts from users who had an insufficient number of sampled tweets ($<25$) to algorithmically classify and who hadn't posted commercial content were classified as Organic.  Due to the high prevalence of hyperlinks included in tweets from promotional accounts, Tweets with URLs whose user had insufficient tweets to classify algorithmically were discarded ( 3.85\% total tweets).    A final list with each tweet classification coding is created by merging the commercial keyword classification with the results from the Human Detection Algorithm.

      \begin{table}[Ht]\caption{Electronic Cigarette Tweet Category Counts}
\begin{tabular}{|ccccc|}
\cline{1-5}
 Year & Total Count & Automated & Organic & Discarded \\
 \hline
  2012 & 107,918 & 85,546 &  13,492 & 8,880\\
  2013 & 426,306  & 339,111 & 76,037 & 11,158 \\
  2014 & 316,424 & 234,972 & 68,698 & 12,754 \\
\hline
\end{tabular}
\end{table}

\section*{Results and Discussion}
  The number of automated, and in particular promotional, tweets vastly overwhelm (80.7\%) the organic (see Figure 2).  The identified automated accounts tweet e-cigarette content  with much higher frequency than the Organic users.  The average number of automated tweets per user was 1.96 with a standard deviation of 35.06 and a max of 14,310.  Average organic posts per user were 1.44 with a standard deviation of 4.01 and max of 356 tweets.   A total of 607,446 Automated Tweets provided a URL (92.09\%).  
   
     \begin{figure}[Ht]
 \centering
  \includegraphics[width=.5\textwidth]{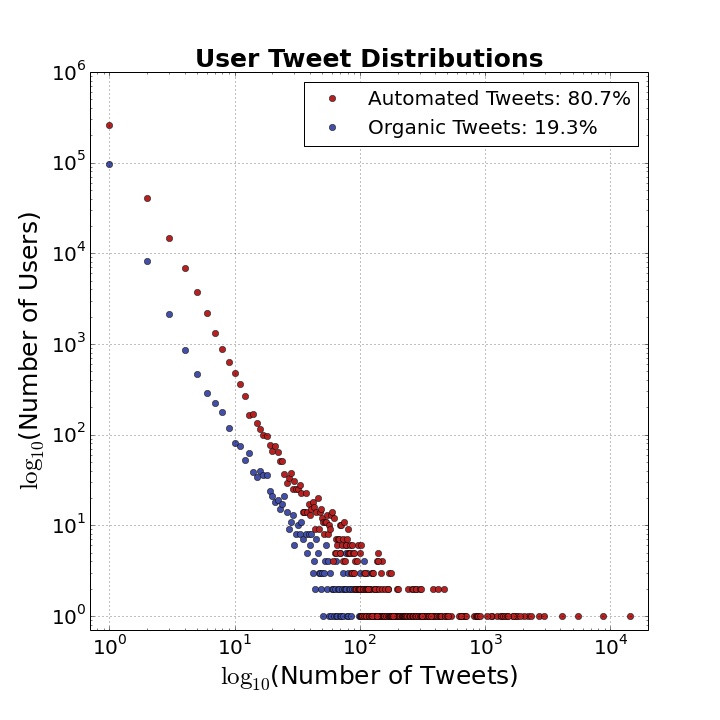}                      
 \caption{ Binned User E-cigarette Keyword Tweet Distribution (2012-2014).  }
\label{fig:UserDist}
 \end{figure}

Frequency WordClouds (see Figure 3) illustrate the most frequently used words by the Automated category.  The size of the text reflects the  ranked word frequencies.  Marketing key words (Free Trial, Brand, Starter Kit, win, Sale) and brand names (V2, Apollo)  are prevalent, illustrating commercial intent.  Many automated tweets also refer to the health benefits of switching to electronic cigarettes (\#EcigsSaveLives), even though they have not been officially approved as such by the Food and Drug Administration, \cite{NYtimesEcig,ashley2007scientific}.  See Table 3 for sub categorical counts of the automated tweets.

   \begin{figure}[Ht]
 \centering
  \includegraphics[scale=.6]{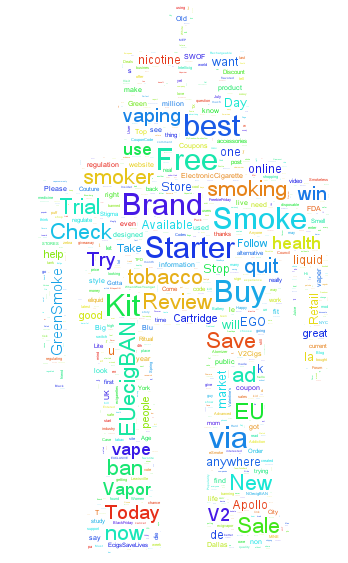}                        
  \caption{ 2013: Automated Tweet Rank-Frequency Word Cloud. High frequency stop words (`of',`the', etc.) are removed from the rank-frequency word distribution. }
\label{fig:RobotCloud}
 \end{figure}

 \subsection*{Tweet Sentiment Analysis}
   Hedonometrics are performed on the organic subset of electronic cigarette tweets to quantify the change in user sentiments over time.   Using the happiness scores of English words from LabMT \cite{dodds2011temporal}, along with its multi-language companion \cite{dodds2014human} the average emotional rating of a corpus is calculated by tallying the appearance of words found  in the intersection of the word-happiness distribution and a given corpus, in this case subsets of tweets.  A weighted arithmetic mean of each word's  frequency, $f_{\textrm{word}}$, and corresponding happiness score, $h_{word}$ for each of the $N$ words in a text yields the average happiness score for the corpus, $\bar{h}_{\textrm{text}}$:     \\
  \vspace{-2mm}
  $$ \bar{h}_{\textrm{text}} = \dfrac{ \sum \limits_{w=1}^{N} f_{\textrm{w}} \cdot h_{\textrm{w}}}{\sum \limits_{w=1}^N f_{\textrm{w}}}$$ 
   
   The average happiness of each word, $h_{avg}$ lies on a 9 point scale: 1 is extremely negative and 9 is extremely positive.  Neutral words ($4\leq h_{avg}\leq 6$), aka `stop words', were removed from the analysis to bolster the emotional signal of each set of tweets.  
   
   \par
    Figure 4 shows that automated electronic cigarette tweets are using very positive language to promote their products.  The average happiness of the Organic tweets are much more stable, and are becoming slightly more negative over time.   Both distributions have a sudden drop in positivity during December 2013, around a debate regarding new e-cigarette legislation by the European Union.  These tweets, labeled \#EuEcigBan, are investigated separately in the next section.  The words that have the largest contributions to changes in sentiments are investigated with Word-shift graphs.
   
      \begin{figure}[Ht]
 \centering
  \includegraphics[width=.5\textwidth]{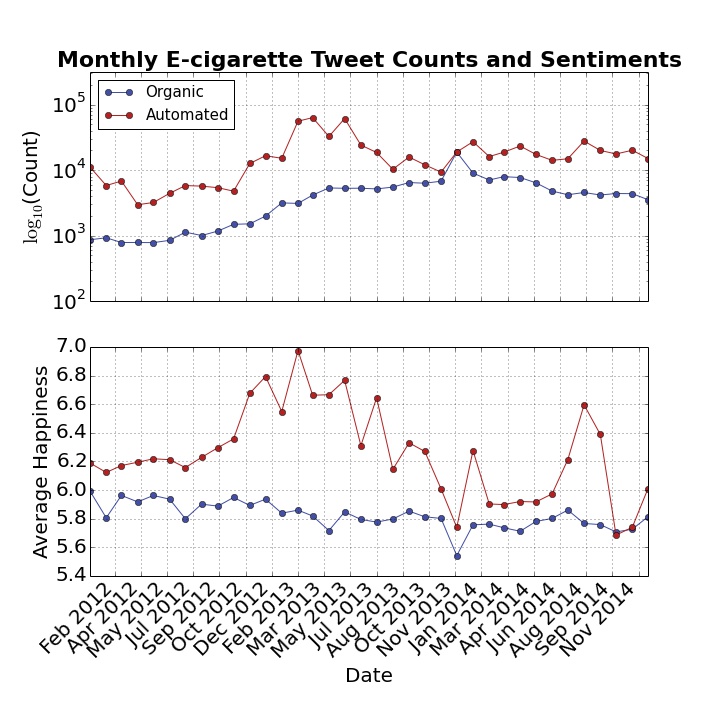}                        
 \caption{ Tweet Frequency and Sentiment Analysis: 2012-2014 }
\label{fig:TweetSentimentTS}
 \end{figure}
 
 Word-shift graphs, introduced in \cite{dodds2011temporal}, illustrate the words causing an emotional shift between two word frequency distributions.  A reference period $(T_{ref})$, creates a basis of the emotional words being used to compare with another period, $(T_{comp})$.  The top 50 words responsible for a happiness shift between the two periods are displayed, along with their contribution to shifting the average happiness of the tweet-set.   The arrows ($\uparrow, \downarrow$) next to a word indicate  an increase or decrease, respectively, of the word's frequency  during the comparison period with respect to the reference period.  The addition and subtraction signs indicate if the word contributes positively or negatively, respectively, to the average happiness score. 
 
 \par

   In Figure 5, below, Word-shift graphs compare the change in Organic sentiments over time, as well as the difference in sentiments between automated and organic tweets.  On the top, the 2013 Organic Tweet distribution is used as a reference to compare sentiments from 2014 Organic Tweets.  December 2013 and January 2014 are removed to dampen the effect of tweets mentioning the \#EUecigBan (see Figure S1).  The average happiness score decreases from 5.84 in 2013 to 5.77 in 2014.   This decrease in the average happiness score is due to a relative increase in the negative words `ban', `tobacco', `doesn't', `drug',  `against', `poison', `tax';  a relative decrease in the positive words `haha', `good', `cool'.  Notably, there is also relatively less usage of the words `quit', `addicted', and an increase in `health', `kids', `juice'.   On the bottom,  Organic tweets from 2013 is the reference distribution to compare Automated tweets from the same year.   Automated tweets are more positive (6.17-6.59 versus 5.84) due to a relative increase in the marketing words `best', `win', `buy', `sale', `health', `discount', etc  and a relative decrease in the negative words `bad', `hate', `stupid', `don't', among others. The words `free' and `trial' are excluded from the graph, since their high frequency and happiness scores distorts the image ($h_{\textrm{avg}}$ increases from 6.17 to 6.59).  
 
   \begin{figure}[Ht]
 \centering
  \includegraphics[scale=.18]{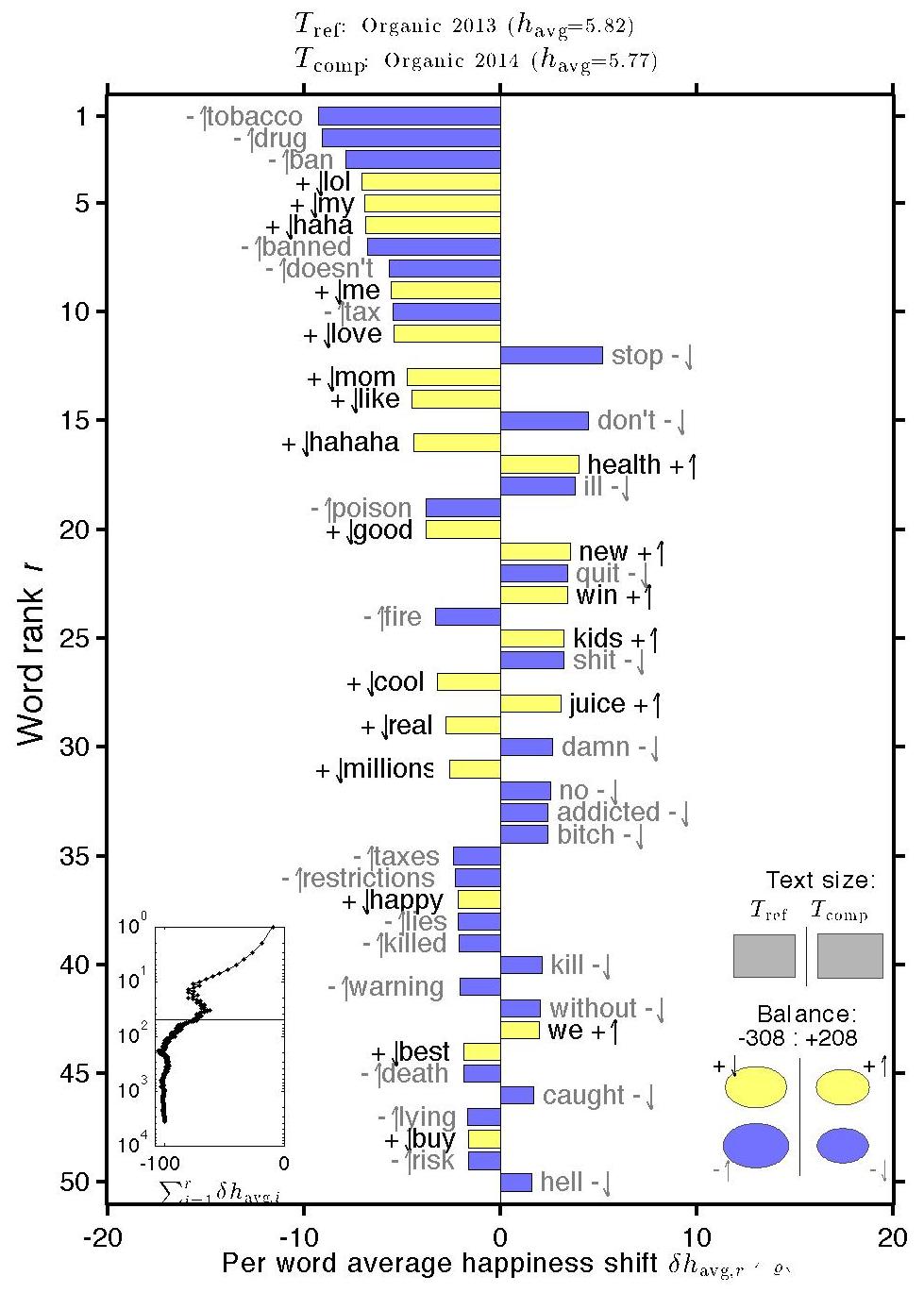}    \\
    \includegraphics[scale=.18]{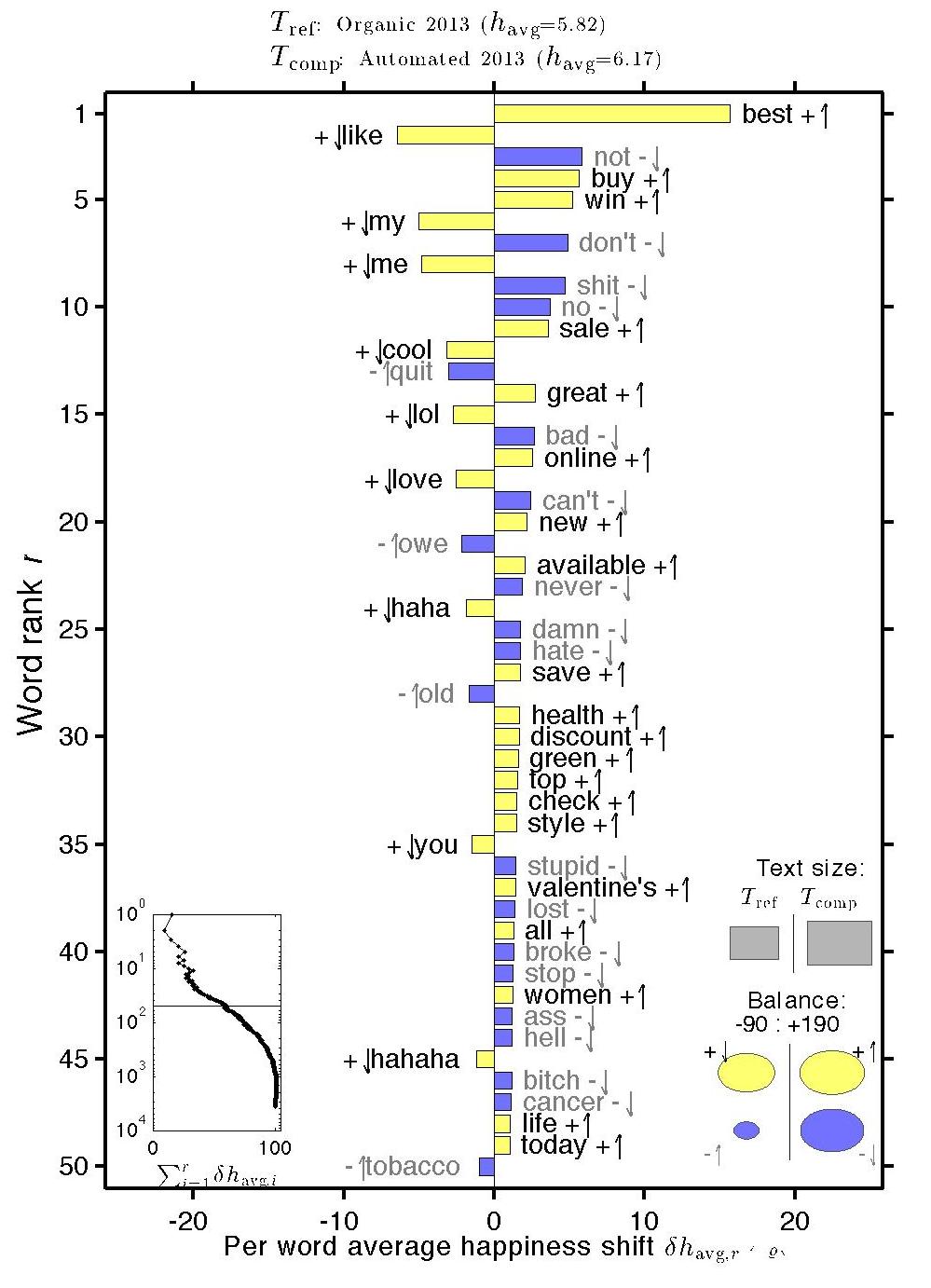}            
 \caption{ \scriptsize{(Top) Organic Tweets from 2013 are the reference distribution to compare sentiments of Organic Tweets from 2014 where we see a negative shift in the calculated average word happiness.   The computed average happiness ($h_{\textrm{avg}}$) decreases from $5.82$ to $5.77$ due to both an increase in the negative words `tobacco', `drug', `ban', `poison', and a decrease in the positive words `love', `like', `haha', `cool' etc.  (Bottom) Organic Tweets from 2013 are the reference distribution to compare Automated Tweets from 2013.   } }
\label{fig:TweetSentimentTS}
 \end{figure}

 \subsection*{Sub-Categorical Tweet Topics}
  \par
	     Pertinent topics related to e-cigarette marketing regulation include  kid-friendly flavors,  smoking cessation claims, and price reduction (including free trials, and starter kits).  Keywords from each of these topics are used to sub-classify the automated tweet set per year, see Table 3 below.  Purely commercial tweets were those with any marketing keywords including: `buy', `save', `coupon(s)', `discount', `price', `cost', `deal', `promo', `money', `sale', `purchase', `offer', `review', `code', `win(ner)', `free', `starter kit(s)', `premium'.  The URL from each tweet was also analyzed for promotional keywords.  Any URL with at least three mentions of the above keywords was enough to classify the tweet as commercial. 
     
     \par
     
     When an individual  on Twitter `follows' another account, posts from these users appear on the `timeline' of the individual.   We quantify the social reach of each of these sub-categorical tweets by counting the total number of accounts' `timelines' who could have been exposed to the advertisement.  To approximate this, we sum the number of followers from each individual's tweets.  The total number of impressions from the commercial category increases from 195.25 million to 951.03 million between 2013 to 2014, even though the total count has dropped from 283k to 149k.  This implies that promotional accounts that are successful in deceiving Twitter's SPAM detector may be gaining many more social links to broadcast their commercial context. 
     
     \par
     
      In order to gauge the accuracy of these sub-categorical tweet topics,  500 tweets were randomly sampled from each category and were evaluated separately by two people to determine the relevance of the tweet to its categorization.  The evaluators had a high level of concordance (84.8\%) and the discrepancies were resolved and merged into a final list.  Sampled tweets were highly relevant per category, the percentage for each is given in Table 3 below. 
     
     \par 
     
      Many automated tweets mentioned using electronic cigarettes as a cessation device, or as a safe alternative.   Over $20,000$ tweets were classified as cessation related, which potentially appeared on over 76.8 million individual's Twitter feed as impressions.  Although electronic cigarettes have not been conclusively authorized as an effective cessation device, \cite{eissenberg2010electronic} has demonstrated the infectiveness of electronic cigarettes to suppress nicotine cravings.  It is also notable that these affiliate marketing accounts are advertising electronic cigarettes as a completely safe alternative to analog tobacco use, contrary to recent studies \cite{sussan2015exposure,MouseEcigTox2015,cameron2014variable,williams2013metal}.   Cessation tweets were tallied using the keywords `quit', `quitting', `stop smoking', `smoke free', `safe', `safer', `safest'.  Many of the purely commercialized tweets mentioned discounts or even free samples.  These Discount tweets were categorized with the keywords `free trial', `coupon(s)', `discount(s)', `save', `sale', `free (e)lectronic (cig)arette'.  Tweets advertising flavors were tallied using the keywords `flavor(s)' and `flavour(s)'.  

\begin{table}[Ht]\caption{Automated Tweet Subcategory  Counts}
\begin{tabular}{|cccccc|}
\cline{1-6}
 Subcategory &  Count  & Percentage & Impressions &  Relevance* & Year\\
\hline
  				    & 53,471   & 62.51\% & 59.74M   & & `12    \\
\textbf{Commercial}      & 283,677   & 83.65\% & 195.25M & 88.4\% & `13    \\
   	   			& 149,333   & 63.55\% & 951.03M & & `14    \\
				
\hline

       & 6,392     &   7.47\%& 8.59M & & `12             \\       
\textbf{Cessation}       & 6,599     &  1.95\%& 25.64M & 90.8\% & `13             \\
       & 8,386     &  3.57\%& 42.72M & & `14             \\
   \hline
       & 26,596     &  31.09\% & 27.02M  & & `12             \\
\textbf{Discount}       & 112,720     &  33.24\% & 38.21M  & 89.8\%& `13           \\
       & 37,735     &  16.06\% & 160.49M  & & `14             \\
       \hline
       & 935     &  1.09\% & 1.73M & & `12            \\
\textbf{Flavor}       & 1,495     &  0.44\% & 2.95M & 81\% & `13           \\
       & 3,833     &  1.63\% & 12.99M  & & `14            \\
\hline

\end{tabular}
*Relevant percentage of 500 randomly sampled tweets
\end{table}

  A noteworthy class of E-cigarette commercial-bots, are those that are masquerading as Organic users to spam pseudo-positive messages towards potential consumers.  These ``cyborgs", as defined in \cite{HumanBotCyborgTwitter,clark2015sifting}, spam a positive message regarding a personal experience.  One class of these automatons are sending contrived testimonies that e-cigarettes have successfully allowed them to quit smoking cigarettes.  These messages are very intentionally structured and tend to swap a few words to appear organic.   These messages also target specific individuals as a more personal form of marketing.  The general tweet structure from a sample cyborg marketing strategy is given below:

\begin{itemize}

\item  \emph{ @USER \{I,We\} \{tried,pursued\}  to \{give up, quit\} smoking . Discovered BRAND electronic cigarettes and quit in \{\#\} weeks. \{Marvelous,Amazing,Terrific\}! URL}
 
 \vspace{1cm}
 
\item \emph{ @USER It's now really easy to \{quit,give up\} smoking (cigarettes).  - these BRAND electronic cigarettes are lots of \{fun,pleasure\}!  URL }

   \vspace{1cm}
  
\item  \emph{@USER electronic cigarettes can assist cigarette smokers to quit, it's well worth the cost URL }
  
   \vspace{1cm}
  
 \item \emph{ @USER It's \{incredible,amazing\} - the (really) \{easy,painless\}  \{answer,method\} to quit cigarette smoking through BRAND electronic cigarettes URL}
  
   \vspace{1cm}
 
 \item \emph{ I managed to quit smoking with these e-cigarettes, I highly recommend them: URL @USER}
 
   \vspace{1cm}
 
 \item \emph{@USER Its \{amazing, extraordinary\} - I (really) quit smoking after \{\#\} yrs thanks to BRAND electronic cigarettes! URL}
  
  \end{itemize}

    \flushleft  
   Using cyborgs to mimic Organic Users for marketing purposes should be analyzed heavily, to gauge their impact and effectiveness on consumers.
   
   \section*{Conclusion}
    Our study has identified an abundance of automated, and in particular, promotional tweets, and consequent organic sentiments.  The collected categorized tweet data from this analysis is available for follow-up analyses into e-cigarette social media marketing campaigns.  Future work can perform a deeper analysis on the URL content, similar to \cite{grana2014smoking}, posted by promotional accounts to get a better sense of the smoking cessation, flavor mentions, and discount prevalence.  We take care not to downplay the well recognized health benefits from smoking cessation including:  decreased risk of coronary artery disease, cerebrovascular disease, peripheral vascular disease, decreased incidence of respiratory symptoms such as cough, wheezing, shortness of breath, decreased incidence of chronic obstructive pulmonary disease, and decreased risk of infertility in women of childbearing age \cite{centers2014health,us2010tobacco,us2004health}. The greatest concern of promotional e-cigarette marketing on Twitter is the risk of enticing  younger generations who otherwise may never have commenced consuming nicotine.  Due to the unknown but unignorable long-term adverse health effects of electronic cigarettes and the alarmingly increased youth consumption of these products, monitoring and  potentially regulating social media commercialization of these products should be immediately relevant to public health and policy agendas.

\subsection{Acknowledgements}

The authors wish to acknowledge the Vermont Advanced Computing Core which provided High Performance Computing resources contributing to the research results. EMC was supported by the UVM Complex Systems Center, PSD was supported by NSF Career Award \# 0846668.  CJ, AK is supported in part by the National Institute of Health (NIH) Research wards R01DA014028 \& R01HD075669, and by the Center of Biomedical Research Excellence Award P20GM103644 from the National Institute of General Medical Sciences.

\clearpage

\setboolean{twocolswitch}{false}

\section*{Supplementary materials}

\includegraphics[scale=.9]{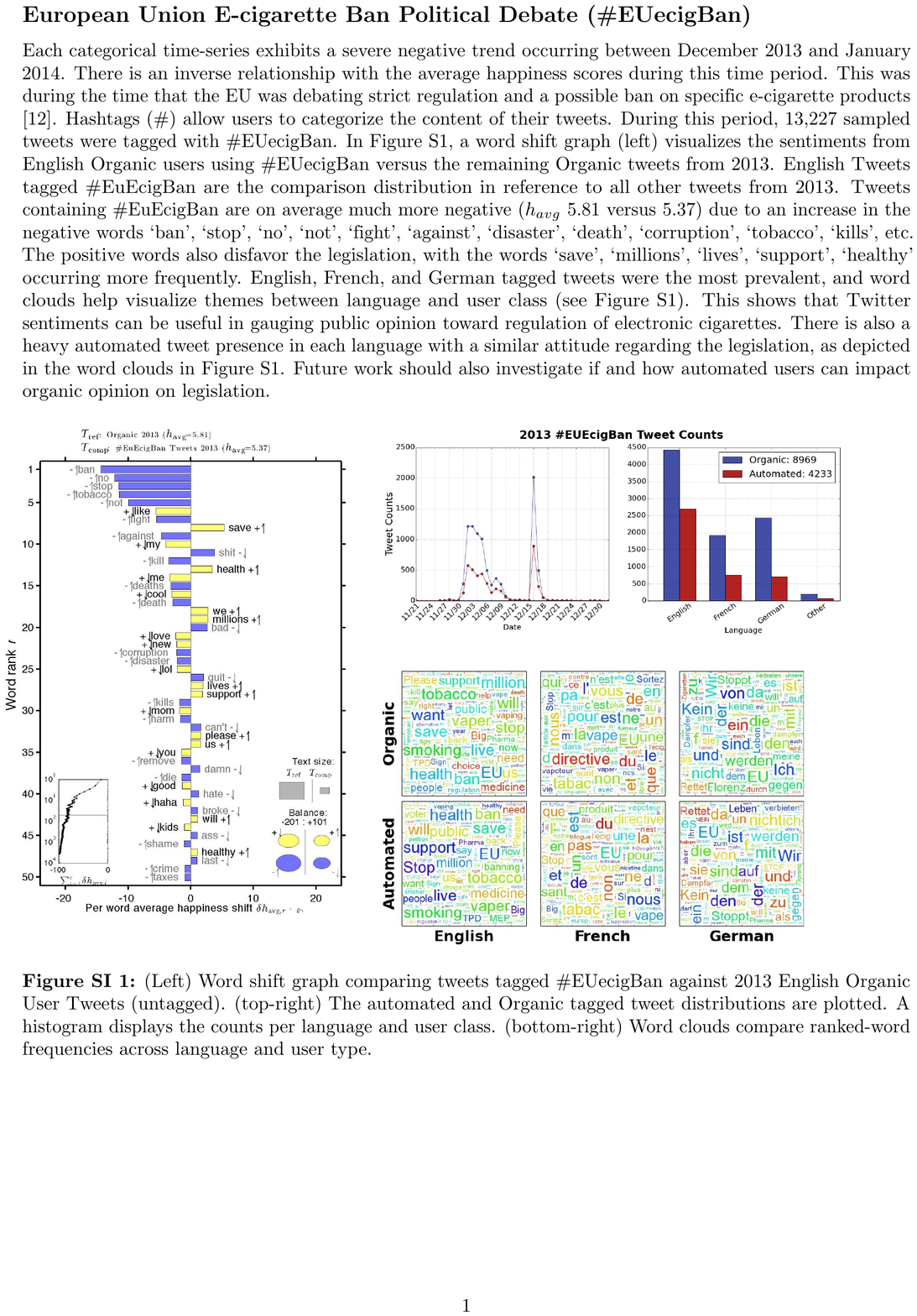}

 \pagebreak


      \begin{table}[H]
\caption{Electronic Cigarette Table of Key Words}
\begin{tabular}{|l|l|}
\cline{1-2}
 Type  & Keywords  \\
 \hline
 & \\
 \textbf{General Twitter Scrape} &  ecig, e cig, e-cig, ecigs, e cigs, e-cigs, e ciggs, \\
     (includes hashtag variants)                                    &  e ciggs, e-ciggs, eciggs, e cigg, ecigg, e-cigarette \\
 &  e cigarette, e cigarettes, e-cigarettes, electronic cigarette  \\ 
 & blucigs, blucig, blu cig, blu cigs, blu ciggs, electronic cigarettes  \\ 
 \hline
 
 \textbf{ Commercial}  & buy, save, coupon, coupons, discount, price, cost, deal, promo, \\
 & money, sale ,purchase, offer,  review, code ,win, winner, \\
 & starter kit, starter kits, premium, \$, kit, \%, sales,voucher, \\
 & brand, free e cigarette, free electronic cigarette, \\
 & free e cig, free ecig \\
 
 \hline
 
 \textbf{ Cessation } & quit, quitting, quits, stop smoking, smoke free, quitter, safe, \\
 				& safest, safer, quitsmoking, give up smoking \\
  
  \hline
  
  \textbf{ Discount } & free trial, free shipping, free sample ,free samples, coupon,\\
                               &  discount, discounts, save, sale, coupons, deal, deals, \\
                               &  free e cigarette, free electronic cigarette, free e cig, free ecig\\ 
                               
         \hline
  
  \textbf{ Flavors* } & flavor, flavour, flavors, flavours, flavored, flavoured  \\
                             & Cherry, Lime, Almond Coconut Bar, Alpine Fresh, Amaretto,\\
			    & Apple Pie (Ala Mode), Banana, Banana Cream, \\
                             &  Banana Graham, Banana Nut Bread ,Banana Pudding,\\
                             & Banana Split, Bavarian Cream, Belgian Waffle \\
                             & Berry Blast, Black Cherry, Black Berry, Black Honey, \\
                             & Blazing Frost, Blueberry,Blueberry Cheesecake, \\
                             & Blueberry Cinnamon Crumble, Blueberry Cotton Candy \\
			    & Blueberry Delight,Brandy, Bubble Gum, Butterscotch \\
			    & Butter Rum, Buttered Popcorn, Cafe Latte, Cake Batter, \\
			    & Candy Cane, Candy Apple, Cantaloupe, Caramel\\
			    & Caramel Cappuccino, Cappuccino,Champagne, \\
			    & Cheesecake, Chocolate Covered Raspberries \\
			    & Cinnamon Coffee Cake, Cinnamon Danish,  \\
			    & Cinnamon Sugar Cookie, Circus Cotton Candy \\
			    & Clove, Coconut, Coconut Candy, Coffee \\
			    &Coffee\&Cream, Cola, Cool, Cotton Candy \\
			    & Cranberry, Crazy Berry, Crazy Chill, Crazy Dew \\
			    & Crazy Freeze, Crazy Grass, Crazy Hump \\
			    & Crazy Pep, Crazy Rainbow, Crazy Watermelon \\
			    & Cream Cheese Frosting, Cream de Menthe \\
			    & Creamy Fruit Smoothie, Cuban Cigar  \\
			    & Cured TobaccoDaquiri, DK-Tab, Double Chocolate \\
			    & Dragon's Blood, Dragon Fruit, Dulce De Leche \\
			    & Egg Nog, English Toffee, Espresso, Extreme Ice \\
			    & Flaming Peach, French Toast, French Vanilla, 
			     French Vanilla Deluxe, Fresh Apple, Fresh-N-Fruity \\
			    & Fudge Brownie, Fruit Rocket, Georgia Peach, Gingerbread, Gummy Candy\\
			    & Goblin Goo, Golden Pineapple, Graham Cracker, Green Apple 
			    Green Tea, Harvest Berry, \\
			    & Hot Chocolate, Hot Cinnamon Candy, Hypnotic, Irish Cream, Hazelnut \\
			    &  Island Getaway, Jamaican Rum, Java Shake, Jungle Juice, Meringue Pie \\
			    & Kentucky Bourbon, Kettle Corn, Khaluah \& Cream, Kiwi, 
			     Lemon Drop, Lemon Lime, Lemon, Mango, \\
			    & Marshmallow, Melon, Menthol, Mint Patty, Milk Chocolate, Munster, N-Mix, N-Mix Menthol, \\
			    & M-Mix Menthol, M-Mix Special Blend, Mocha, Mojito, Mummy Mint \\
			    & NY Cheesecake,  Orange Creamsicle, P-Mix, P-Mix Menthol, Papaya Passion Fruit, \\
			    & Peanut Butter, Peanut Buttercup, Honey Dew Melon, Margarita, M-Mix, Orange Cognac\\
			    \hline		    
\end{tabular}
*Flavors compiled from \url{https://crazyvapors.com/e-liquid-flavor-list/}   Keywords other than `General Twitter Scrape' were applied to categorize automated account tweets
\end{table}

\end{document}